 \documentclass[bibyear]{aa}                
\usepackage{graphicx}
\usepackage{txfonts}
\begin{document}

\title{Mapping possible non-Gaussianity in the Planck maps}


\author{
A. Bernui\inst{1}
\and
M. J. Rebou\c{c}as\inst{2}
          }

\institute{Observat\'orio Nacional, Rua General Jos\'e Cristino 77,
               20921-400 Rio de Janeiro -- RJ, Brazil \\
               \email{bernui@on.br}
\and
              Centro Brasileiro de Pesquisas F\'{\i}sicas,
              Rua Dr.\ Xavier Sigaud 150,
              22290-180 Rio de Janeiro -- RJ, Brazil \\
              \email{reboucas@cbpf.br}
              }


\date{ }


\abstract
{The study of the non-Gaussianity of the temperature fluctuations of cosmic background radiation
(CMB) can be used to break the degeneracy between the inflationary models and to test alternative
scenarios of the early universe. However, there are several sources of non-Gaussian contaminants
in the CMB data, which make a convincing extraction of primordial non-Gaussianity into an ambitious
observational and statistical enterprise.
It is conceivable that no single statistical estimator can be sensitive to all forms and levels of
non-Gaussianity that may be present in observed CMB data.
In recent works a statistical procedure based upon the calculation of the skewness and kurtosis
of the patches of CMB sky sphere has been proposed and used to find out significant large-angle
deviation from Gaussianity in the foreground-reduced WMAP maps.
}
{Here we address the question of how previous recent analyses of Gaussianity of WMAP maps
are modified if the nearly full-sky foreground-cleaned  Planck maps are used, therefore extending
and complementing such an examination in several regards.
}
{Once the foregrounds are cleaned through different component separation procedures, each
of the resulting Planck maps is then tested for Gaussianity.
We determine quantitatively the effects for Gaussianity when masking the  foreground-cleaned
Planck maps with the  {\sc inpmask}, {\sc valmask}, and U73 Planck masks.
}
{We show that although the foreground-cleaned Planck maps present significant deviation from
Gaussianity of different degrees when the less severe {\sc inpmask} and {\sc valmask} are used, they
become consistent with Gaussianity as detected by our indicator $S$ when masked with the union
U73 mask.
A slightly smaller consistency with Gaussianity is found when the $K$ indicator is employed,
which seems to be associated with the large-angle anomalies reported by the Planck team.
Finally, we examine the robustness of the Gaussianity analyses with respect to the real pixel's noise
as given by the Planck team, and show that no appreciable changes arise when it is incorporated into
the maps.
The results of our analyses  provide important information concerning Gaussianity of the
foreground-cleaned Planck maps when diverse cut-sky masks are used.
}
{}


   \maketitle
%

\section{Introduction} \label{introduction}

The most general spacetime geometry consistent with the principle of spatial homogeneity and
isotropy and the existence of a cosmic time is the Friedmann-Lema\^{\i}tre-Robertson-Walker
(FLRW) metric.
Within the FLRW approach to cosmological modeling in the framework of general relativity,
the additional suggestion that the Universe underwent a period of rapid accelerating expansion
(Starobinsky~\cite{Starobinsky79},~\cite{Starobinsky82}; Kasanas~\cite{Kasanas};
Sato~\cite{Sato}; Guth~\cite{Guth}; Linde~\cite{Linde82}; Albrecht \& Steinhardt~\cite{Albrecht})
has become an essential building block of the standard cosmological model.
Besides solving the so-called horizon and flatness problems that come out in the FLRW model, the
cosmological inflation provides a mechanism for the production of the primordial density fluctuations,
which seeded the observed cosmic microwave background (CMB) anisotropies and the formation
of large structures we observe in the Universe.

However, there are a great number of inflationary models, among which the simplest ones are
based on a slowly-rolling single scalar field (see, for example, Bassett et al.~\cite{Bassett};
Linde~\cite{Linde08}).
An important prediction of these simple models is that, regardless of the form of the kinetic term,
the potential, or the initial vacuum state, they can generate only tiny primordial
non-Gaussianity (Creminelli \& Zaldarriaga~\cite{CreminelliZaldarriaga2004};
Komatsu~\cite{Komatsu-2010}).
Thus, although convincing detection of a fairly large primordial non-Gaussianity would not
exclude all inflationary models, it would rule out the entire class of simple models (see, e.g.,
Bartolo et al.~\cite{Bartolo2004}; Komatsu et al.~\cite{WMAP-Komatsu09};
Bassett et al.~\cite{Bassett}).
Moreover, robust stringent constraints on primordial non-Gaussianity would rule out alternative
models of the early universe (see, for example, Koyama et al.~\cite{Koyama}; Buchbinder et
al.~\cite{Buchbinder}, Lehners \& Steinhardt~\cite{Lehners}; Cai et al.~\cite{Cai09a},~\cite{Cai09b}).
In this way, the study of  primordial non-Gaussianity offers an important window into the physics of
the early universe.

Gravitational waves generated by inflation provide another crucial window since they induce local
quadrupole anisotropies in the radiation bombarding free electrons within last-scattering surface,
inducing polarization in the scattered CMB photons.
This polarized radiation includes the B-mode component that cannot be generated by density
perturbations.
Thus, detection of primordial B-mode polarization of CMB provides a unique confirmation of a
crucial prediction of the simple inflationary models.
Report of the detection of this B-mode polarization  in the CMB has recently
been made by BICEP Collaboration (Ade et al.~\cite{BICEP2}). But since the BICEP2's
announcement, concerns about the impact of our Galaxy's dust contribution to the BICEP2 results
have been raised  (see, for example, Flauger et al.~\cite{Flauger2014} and
Adam et al.~\cite{Planck-XXX}).

In the study of deviation from Gaussianity of the CMB temperature fluctuations data, one
is particularly interested in the primordial component. However, there are several sources
of non-Gaussianity in the observed CMB data that do not have a primordial origin, including
contributions from residual diffuse foreground emissions (Bennett et al.~\cite{Bennett-etal-2003};
Leach et al.~\cite{Leach-etal-2008}; Rassat et al.~\cite{Rassat}), unresolved point sources
(Komatsu et al.~\cite{Komatsu03}), and systematic instrumental effects
(Donzelli et al.~\cite{Donzelli_etal-2009}, Su et al.~\cite{Su-etal2010}), secondary CMB anisotropies,
such as the Sunyaev-Zel'dovich effect (Zel'dovich \& Sunyaev~\cite{SZ-1969}; Novaes \&
Wuensche~\cite{Novaes}), gravitational lensing (see Lewis \& Challinor~\cite{Lewis-Challinor2006}
for a review), and the integrated Sachs-Wolfe effect (Rees \&
Sciama~\cite{Rees-Sciama_1968}).\footnote{The combination of the ISW and gravitational lensing
produces the dominant contamination to the non-Gaussianity of local type (Goldberg \&
Spergel~\cite{Goldberg_Spergel-1999}).}
These contaminant non-Gaussian sources make a reliable extraction of primordial
non-Gaussianity into a challenging observational and statistical enterprise.

Since there is no unique signature of non-Gaussianity, it is conceivable that no single statistical
estimator can be sensitive to all sources of non-Gaussianity that may be present in observed
CMB data. Furthermore, different statistical tools can provide valuable complementary information
about different features of non-Gaussianity.
Thus, is it important to test CMB data for deviations from Gaussianity by employing different
statistical estimators  to examine the non-Gaussian signals in the Planck CMB data and, possibly,
to shed light on their source.
The Planck collaboration has analyzed non-Gaussianity  by using two classes of statistical tools
(Planck Collaboration I, XXIII, XXIV~\cite{Planck-I}).
In the first, the  parametric optimal analyses were carried out to constrain the  model-dependent
amplitude parameter $f_{\rm NL}$ for the local, equilateral, and orthogonal bispectrum types.
An important consistency with Gaussianity, at $1\sigma$ confidence level, has been found by
narrowing considerably the Wilkinson Microwave Anisotropy Probe (WMAP) interval of  $f_{\rm NL}$
for the three types (shapes) of non-Gaussianity (Planck Collaboration XXIV~\cite{Planck-XXIV}).
The second class contains a number of model-independent tools, including the spherical Mexican-hat
wavelet (Mart\'{\i}nez-Gonz\'alez~\cite{Wavelet-2002}),
Minkowski functionals (Minkowski~\cite{Minkowski}; Gott et al.~\cite{Gott}; Komatsu et
al.~\cite{Komatsu03}; Komatsu et al.~\cite{KomatsuWMAP5}; Eriksen et al.~\cite{Eriksen};
De Troia et al.~\cite{Troia}; Curto et al.~\cite{Curto}; Novaes et al.~\cite{Novaes14}),
and the surrogate statistical technique (R\"ath et al.~\cite{Rath09}; R\"ath et al.~\cite{Rath11};
Modest et al.~\cite{Modest}).
Little evidence of non-Gaussianity was obtained through most of these former indicators, but there
are non-Gaussianity tools, such as the surrogate map technique, which suggests significant
deviation from Gaussianity in CMB Planck data (Planck Collaboration XXIII~\cite{Planck-XXIII}).

One of the simplest Gaussianity tests of a  CMB map can be made by computing  the skewness and
kurtosis from the whole set of accurate temperature fluctuations values.
This procedure would furnish two dimensionless overall numbers for describing Gaussianity,
which could be compared, e.g., with the values of the statistical moments calculated from Gaussian
simulated maps.
The Planck collaboration has employed this procedure by calculating a sample of
values of the skewness and kurtosis from the four foreground-cleaned U73-masked Planck maps,
and made a comparison  with the averaged values for the statistical moments obtained from
simulated maps (Planck collaboration XXIII~\cite{Planck-XXIII}).\footnote{Different values for
$N_{\rm \sc side}$ were used by Planck team to show the robustness of their results with respect to
the angular resolution of CMB maps.}

However, one can go a step further and obtain a large number of values of the skewness and kurtosis
along with directional and angular-scale information about large-angle non-Gaussianity if one divides
the CMB sphere $\mathbb{S}^2$ into a number $j$ (say) of uniformly distributed spherical
patches of equal area that  cover $\mathbb{S}^2 $.
We then calculate the skewness, $S_j$, and the kurtosis, $K_j$, for each patch $j=1, \ldots ,n$.
The union of values $S_j$ and $K_j$ can thus be used to define two discrete functions
$S(\theta,\phi)$ and $K(\theta,\phi)$ in such way that $S(\theta_j,\phi_j)= S_j$ and
$K(\theta_j,\phi_j)=K_j$ for every $j=1, \ldots ,n$.  This is a constructive way of
defining two discrete functions from any given CMB maps, which provide local measurements
of the non-Gaussianity as a function of angular coordinates $(\theta,\phi)$.
The Mollweide projections of $S(\theta_j,\phi_j)$ and $K(\theta_j,\phi_j)$
are skewness and kurtosis maps, whose power spectra can be  
used to study two large-angle deviation from Gaussianity. 

This statistical procedure based upon calculating the skewness and kurtosis of the patches
of CMB sky sphere has recently been proposed in Bernui \& Rebou\c{c}as (\cite{BR2009}) and used
to examine deviation from Gaussianity in simulated maps (Bernui \& Rebou\c{c}as~\cite{BR2012}),
as well as in the foreground-reduced WMAP maps (Bernui \& Rebou\c{c}as~\cite{BR2010};
see also Zhao~\cite{Zhao}; Bernui et al.~\cite{BMRT}).
We note that these estimators capture directional information of non-Gaussianity and are useful in
the presence of anisotropic signals, as foregrounds, for example.
A pertinent question that arises here is how the analysis of Gaussianity made with WMAP maps is
modified if the foreground-cleaned maps recently released by Planck collaboration are used.
Our primary objective in this paper is to address this question  by extending and complementing the
analyses of Bernui \& Rebou\c{c}as~(\cite{BR2010}) in four different ways.
First, we use the same statistical indicators to carry out a new analysis of Gaussianity of the available
nearly full-sky foreground-cleaned Planck maps, which have been produced from CMB Planck
observations in the  nine frequency bands between $30$ GHz and $875$ GHz through different
component separation techniques.
Second, since the foregrounds are cleaned through different component separation procedures, each
of the resulting foreground-cleaned Planck maps is thus tested for Gaussianity.
Then, we make a quantitative analysis of the effects of different component separation cleaning
methods in the Gaussianity and quantify  the level of non-Gaussianity for each foreground-cleaned
Planck map.
Third, we quantitatively study the effects for the analyses of Gaussianity of masking the
foreground-cleaned Planck maps with their {\sc inpmask}, {\sc valmask}, and U73 masks.
Fourth, we use Planck pixel's noise maps to examine the robustness of the Gaussianity analyses with
respect to the  pixel's noise as given by Planck team.
These analyses of deviation from Gaussianity in the Planck maps provide important information about
the suitability of the Planck maps as Gaussian reconstructions of the CMB sky,
when diverse cut-sky masks are used.

The structure of the paper is as follows.
In Sec.~\ref{Planck-maps} we briefly present the Planck foreground-cleaned maps and the masks we
use in this paper.
In Sec.~\ref{Indicators} we present our non-Gausssianity statistical indicators and  the associated
skewness and kurtosis maps.
Section~\ref{NG-Planck-maps} contains the results of our analyses withthe foreground-cleaned Planck
maps,  and f\/inally in Sec.~\ref{Conclusions} we present the summary of our main results and conclusions.

\section{Foreground-cleaned maps and masks} \label{Planck-maps}

Planck satellite has scanned the entire sky in nine frequency bands centered at $30$, $44$, $70$,
$100$, $143$, $217$, $353$, $545$, and $857$ GHz with angular resolution varying from
$\sim 30$ to $\sim 5$ arcminutes.
These observations have allowed Planck collaboration to reconstruct the CMB temperature
fluctuations over nearly the full sky.
To this end, they used four different component separation techniques, aimed at removing the
contaminants, including emissions from the Galaxy, which are present on large angular scales,
and extragalactic foregrounds (and compact sources), which are dominant on small scales.
These techniques are based on two different methodological approaches.
In the first, only minimal assumptions about the foregrounds are made, and it is sought
to minimize the variance of CMB temperature fluctuations for a determined blackbody spectrum,
while the second essentially relies on parametric modeling of the foreground.
By using the component separation techniques Planck collaboration produced four and
released three nearly full-sky foreground-cleaned CMB maps (Planck Collaboration
XII~\cite{Planck-XII}), which are called {\sc smica} (spectral matching independent component
analysis, Cardoso et al.~\cite{Cardoso08}), {\sc nilc} (needlet internal linear combination,
Delabrouille et al.~\cite{Delabrouille09}), {\sc sevem} (spectral estimation via expectation
maximization, Fern\'andez-Cobos et al.~\cite{Fernandez-Cobos12}).

Each CMB Planck map is accompanied by its own confidence mask or validation mask ({\sc valmask},
for short), outside which one has the pixels  whose values of the temperature fluctuations are expected
to be statistically robust and to have a negligible level of foreground contamination.
Since the component separation techniques handle the data differently, the corresponding
{\sc valmask}s are different for each foreground-cleaning map-making procedure.
Besides the {\sc valmask}s, the {\sc smica} and {\sc nilc} maps were released with their corresponding
minimal masks, called {\sc inpmask}s.

The allowed sky fraction for a given mask, i.e. the fraction of unmasked data pixels, is denoted by
$f_{\mbox{sky}}$.   
In Table~\ref{table1} we collect the $f_{\mbox{sky}}$ values, which give the fraction of
unmasked pixels for the three foreground-cleaned Planck maps that we used in
the statistical analyses of this article.
Thus, for example, for the mask called U73, which is the union of the confidence masks, one has
$f_{\mbox{sky}}=0.73$, since it allows only $73\%$ of the CMB sky.

\begin{table}[h]
\begin{center}
\begin{tabular}{lc} 
\hline \hline
 Planck mask \ \ & \ \ \ \ \ \ \ \ \ $f_{\mbox{sky}}$ \\ 
\hline
{\sc smica}~--~{\sc inpmask}   & \ \ \ \ \ \ \ \ \ 0.97   \\
{\sc smica}~--~{\sc valmask}   & \ \ \ \ \ \ \ \ \ 0.89   \\
{\sc nilc}~--~{\sc inpmask}    & \ \ \ \ \ \ \ \ \ 0.97   \\
{\sc nilc}~--~{\sc valmask}    & \ \ \ \ \ \ \ \ \ 0.93   \\
{\sc sevem}~--~{\sc valmask}   & \ \ \ \ \ \ \ \ \ 0.76   \\
U73~--~ {\sc union-mask}       & \ \ \ \ \ \ \ \ \ 0.73   \\
\hline \hline
\end{tabular}
\end{center}
\caption {Sky fraction values $f_{\mbox{sky}}$, which give the fraction of
unmasked pixels for the three foreground-cleaned Planck maps.} \label{table1}
\end{table}

In addition to the above-mentioned foreground-cleaned maps and associated masks,the Planck
team has produced and released noise maps that contain an estimate of the pixel's noise,
particularly from the non-uniform strategy for scanning the CMB sky and noise of instrumental
nature.

\section{Non-Gaussianity statistical procedures and indicators} \label{Indicators}

In this section we describe two non-Gaussian statistical indicators and the procedure
for calculating their associated maps from CMB temperature fluctuations maps.
The procedure described here is used in the following sections to investigate
large-angle deviation from Gaussianity in the foreground-cleaned Planck maps.

Our primary purpose is to give a procedure for defining, from a given CMB map,
two discrete functions, $S(\theta_j, \phi_j)$ and $K(\theta_j, \phi_j)$, on the
two-sphere $\mathbb{S}^2$ that measure deviation from Gaussianity in the directions
given by $(\theta_j, \phi_j)$, where $(\theta, \phi)$ are spherical coordinates and
$j=1, \ldots , n$ for a chosen integer $n$.
The Mollweide projections of these two functions give what we call $S$ and $K$ maps,
whose large scale features are employed to investigate large-angle deviation from Gaussianity.

To construct such functions, a first important ingredient is that deviation from Gaussianity of the
CMB temperature fluctuations data can be measured by calculating the skewness $S$ and the
kurtosis $K$ from temperature fluctuations data in patches, which are subsets of the CMB sky-sphere
containing a number of pixels with their corresponding data values.

The second essential ingredient in the practical construction of $S(\theta_j, \phi_j)$  and
$K(\theta_j, \phi_j)$ is  the choice of these patches, to calculate $S$ and $K$ from the data therein.
Here we chose these patches to be spherical caps (calottes) of aperture $\gamma$, centered on
homogeneously distributed points on the CMB two-sphere
$\mathbb{S}^2$.
This choice was made in Refs. Bernui \& Rebou\c{c}as (\cite{BR2009},~\cite{BR2010},
and~\cite{BR2012}).

The above-outlined points of the procedure to define the discrete functions
$S(\theta_j, \phi_j)$ and  $K(\theta_j, \phi_j)$ from a given CMB  map can
be formalized through the following steps:
\begin{enumerate}
\item[{\bf i.}]
Take on the CMB two-sphere $\mathbb{S}^2$ a discrete set of  $n$ points
homogeneously distributed as the centers of spherical caps of a given aperture
$\gamma$ (say);
\item[{\bf ii.}]
Using the values of the temperature fluctuations of a given CMB map, calculate,
for each cap of the above item {\bf i},  the skewness and kurtosis given,
respectively, by
\begin{eqnarray}
&&S_j   =  \frac{1}{N_{\rm p} \,\sigma^3_{\!j} } \sum_{i=1}^{N_{\rm p}}
\left(\, T_i\, - \overline{T_j} \,\right)^3 \,, \label{S-Def}\\
&&K_j   =  \frac{1}{N_{\rm p} \,\sigma^4_{\!j} } \sum_{i=1}^{N_{\rm p}}
\left(\,  T_i\, - \overline{T_j} \,\right)^4 - 3 \label{K-Def} \,,
\end{eqnarray}
for $j = 1, 2, \cdots, n$, and where $N_{\rm p}$ is the number of pixels in the
$j^{\,\rm{th}}$ cap; $T_i$ is the temperature at the $i^{\,\rm{th}}$ pixel in the
$j^{\,\rm{th}}$ cap; $\overline{T_i}$ the CMB mean temperature in the $j^{\,\rm{th}}$
cap; and $\sigma_j$ the standard deviation for the temperature fluctuations data of the
$j^{\,\rm{th}}$ cap.

Clearly, the values $S_j$ and $K_j$ obtained in this way for each cap give a measure
of non-Gaussianity in the direction $(\theta_j, \phi_j)$ of the center of each
$j^{\,\rm{th}}$ cap.
\item[{\bf iii.}]
Finally, use the union of all calculated values $S_j$  to define a discrete function
$S(\theta,\phi)$ on the two-sphere  such that $S(\theta_j,\phi_j)= S_j$ for $j=1, \ldots ,n$.
Similarly use the values $K_j$ to define another discrete function $K(\theta,\phi)$
on $\mathbb{S}^2$ through the equation $K(\theta_j,\phi_j)=K_j$ for $j=1, \ldots ,n$.
The Mollweide projection of these functions constitutes skewness and kurtosis maps, hereafter
denoted as $S$-maps and $K$-maps, respectively, which we use to investigate the deviation
from Gaussianity as a function of the angular coordinates $(\theta,\phi)$.
These maps give a  directional and geographical distribution of skewness and kurtosis
values calculated from a given CMB input map.
\end{enumerate}

Now, the discrete functions $S(\theta_j, \phi_j)$ and $K(\theta_j, \phi_j)$ defined
on the two-sphere $\mathbb{S}^2$ can be expanded into their spherical harmonics and
their angular power spectra can be calculated from the corresponding coefficients.
Thus, for example, for   $S(\theta_j, \phi_j)$
one has
\begin{equation}
S(\theta_j, \phi_j) = \sum_{\ell=0}^\infty \sum_{m=-\ell}^{\ell}
a_{\ell m} \,Y_{\ell m} (\theta,\phi) \; ,
\end{equation}
and  the corresponding angular power spectrum
\begin{equation}
S_{\ell} = \frac{1}{2\ell+1} \sum_m |a_{\ell m}|^2 \; .
\end{equation}
Clearly, one can similarly expand the kurtosis function $K(\theta_j, \phi_j)$  and
calculate its angular power spectrum $K_{\ell}$.
Since we are interested in the large angular scale information, we restrict our analyses
to the low $\ell$ spectra, i.e., $\,\ell=1,\,\,\cdots,10\,$.

In the next sections we use the power spectra $S_\ell$ and $K_\ell$ to assess possible departures
from Gaussianity of the {\sc smica}, {\sc nilc}, and {\sc sevem} foreground-cleaned Planck maps
and to calculate the statistical significance of potential deviation from Gaussianity by comparison
with the corresponding power spectra calculated from input Gaussian maps.

\begin{figure*}[thb!] 
\begin{center}
\includegraphics[width=4.2cm,height=7.0cm,angle=90]{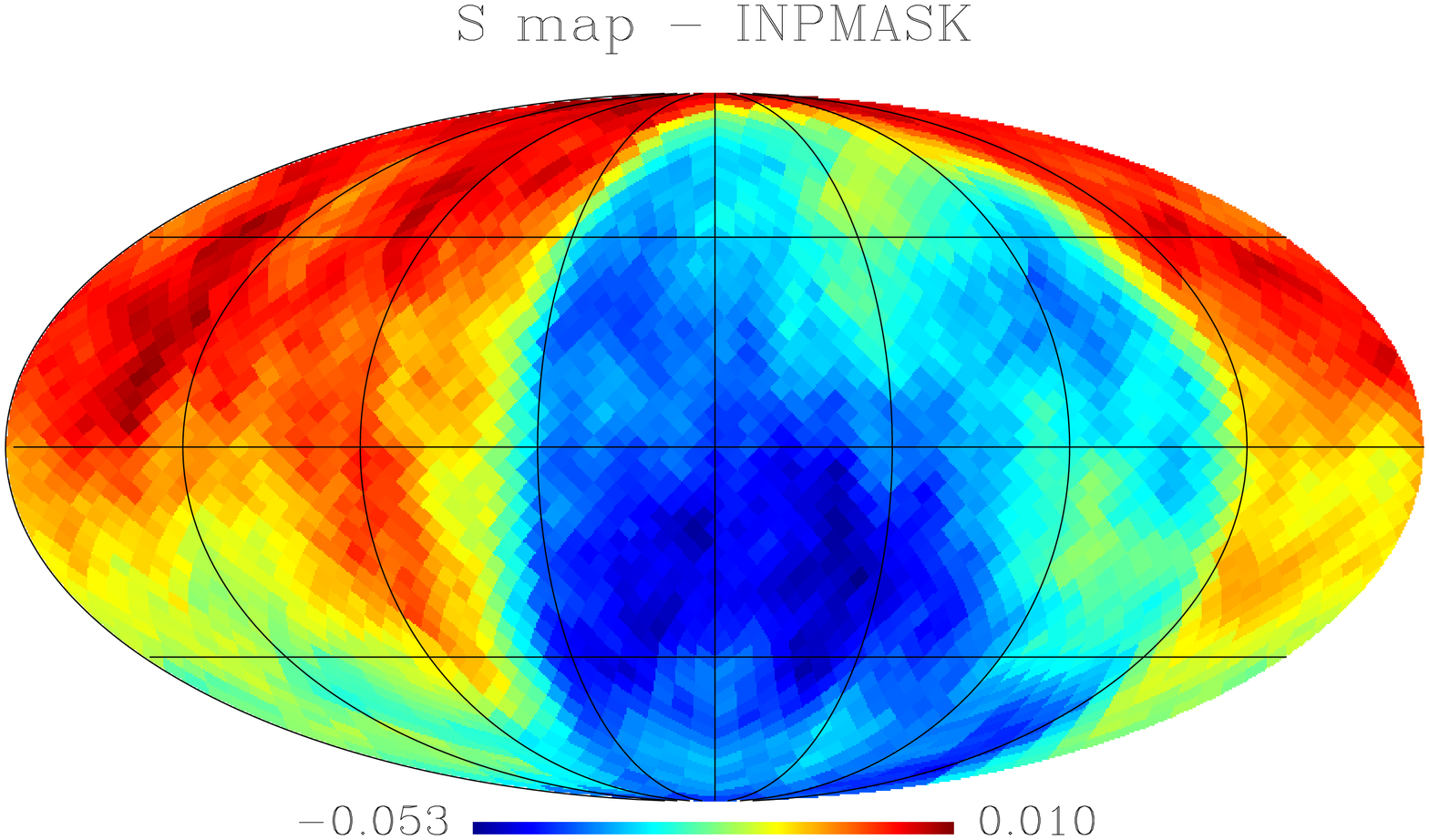}
\hspace{3mm}
\includegraphics[width=4.2cm,height=7.0cm,angle=90]{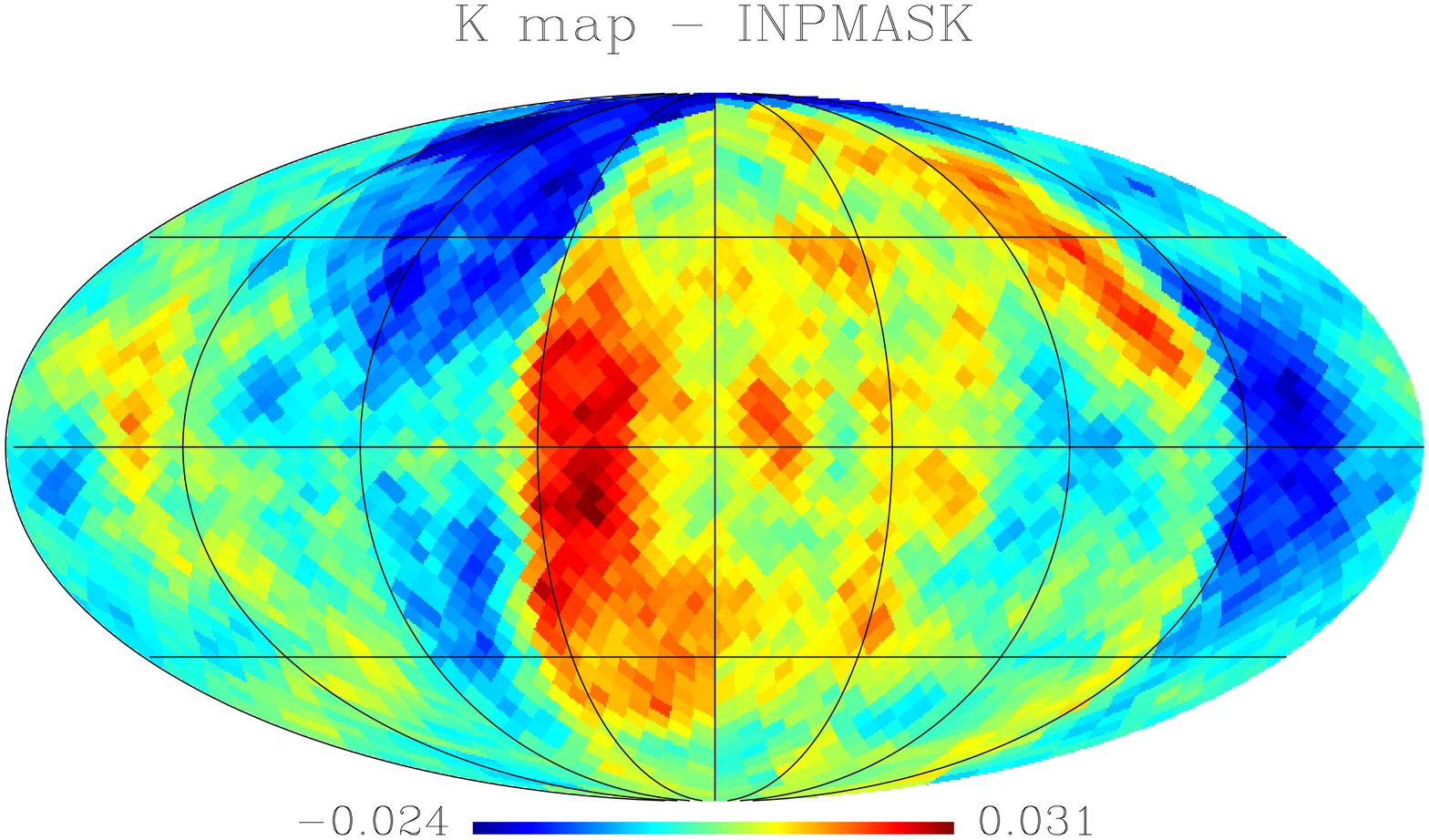}
\vspace{2mm} \vspace{2mm} 
\includegraphics[width=4.2cm,height=7.0cm,angle=90]{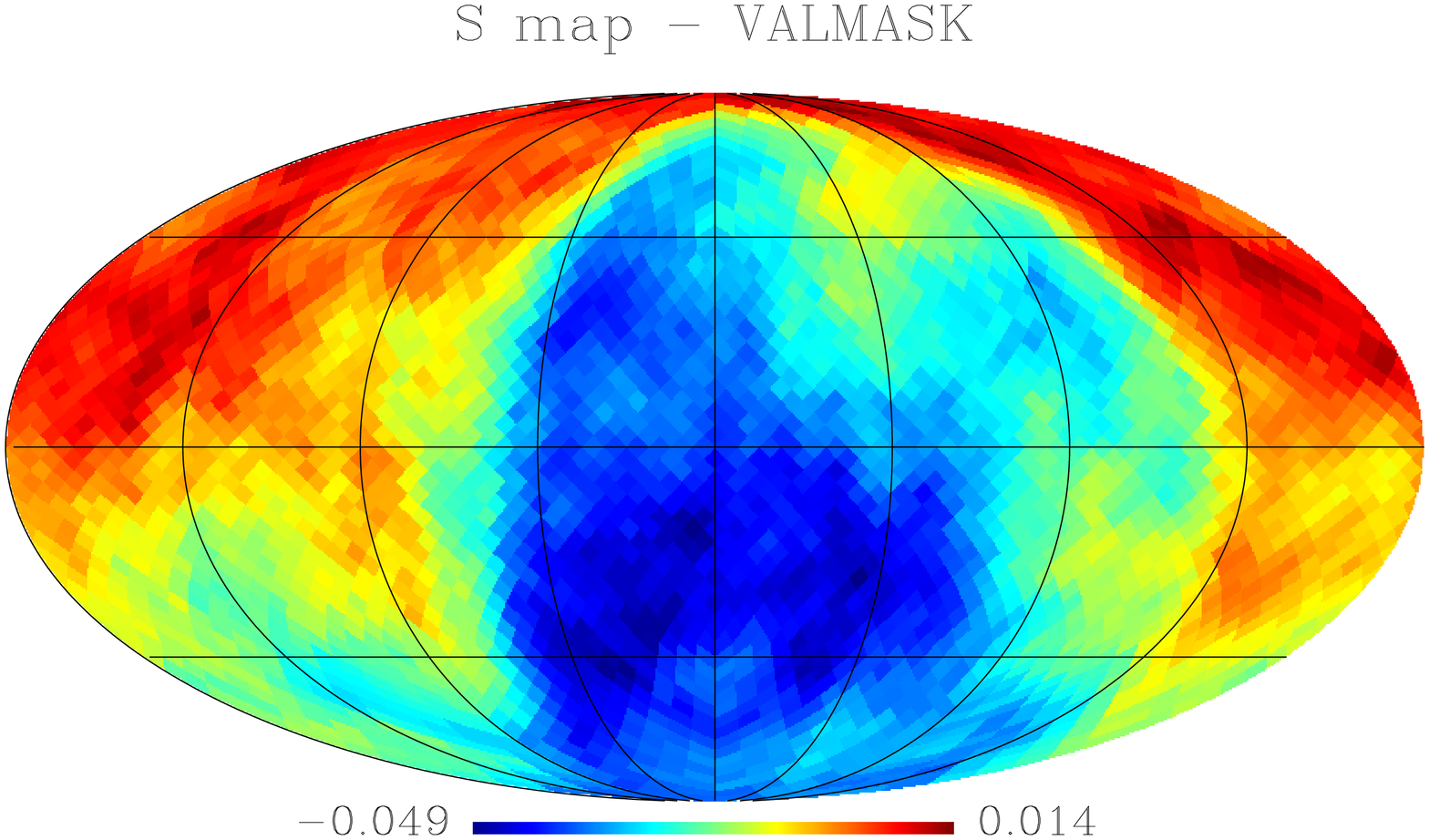}
\hspace{3mm}
\includegraphics[width=4.2cm,height=7.0cm,angle=90]{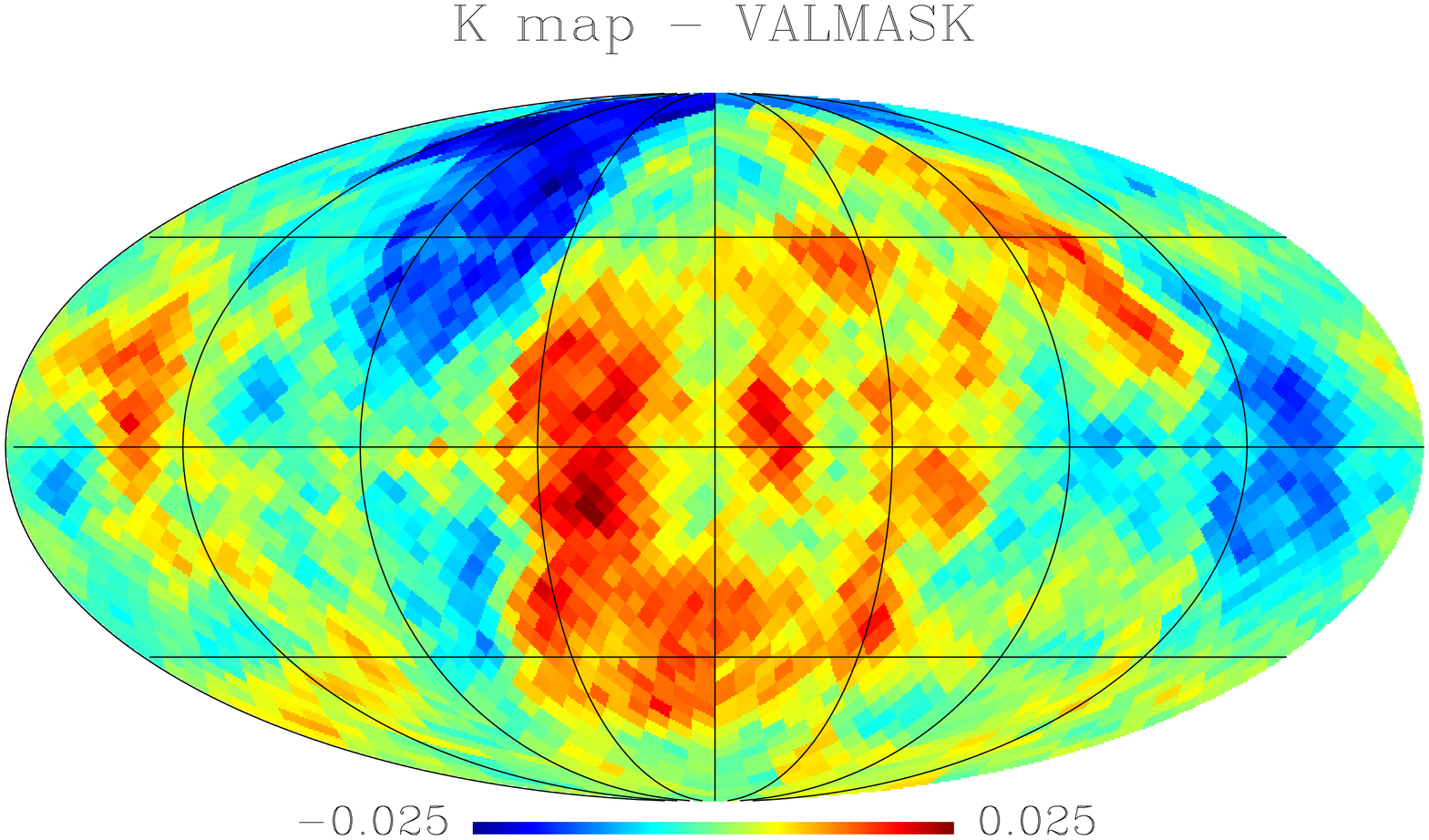}
\caption{$S$ (left) and $K$ (right) maps calculated from the {\sc smica} map
with $N_\text{side}=512$. The {\sc inpmask} was used for generating the maps in the
first row, while in the second row the {\sc valmask} was employed to generate
the maps. \label{Fig1} }
\end{center}
\end{figure*}

\section{Non-Gaussianity in the Planck maps} \label{NG-Planck-maps}

The use of the non-Gaussian statistical procedure described in Section~\ref{Indicators} requires not
only the choice of the CMB input maps, but also specification of some important quantities that are
required to carry out the computational routine.
In this way, to minimize the statistical noise in the calculation of the functions $S(\theta_j, \phi_j)$
and $K(\theta_j, \phi_j)$ from Planck input maps, we have scanned the celestial sphere with
spherical caps of aperture $\gamma = 90^{\circ}$, centered at $3\,072$ points homogeneously
distributed on the two-sphere $\mathbb{S}^2$ and generated by using  {\sc healp}ix package
(G\'orski et al.~\cite{Gorski-et-al-2005}).
This robust choice of quantities has been established in Bernui \& Rebou\c{c}as
(\cite{BR2009},~\cite{BR2010}), and thoroughly tested recently by using simulated CMB maps in
Bernui \& Rebou\c{c}as (\cite{BR2012}).

\begin{figure*}[t!] 
\begin{center}
\includegraphics[width=6.5cm,height=5.cm]{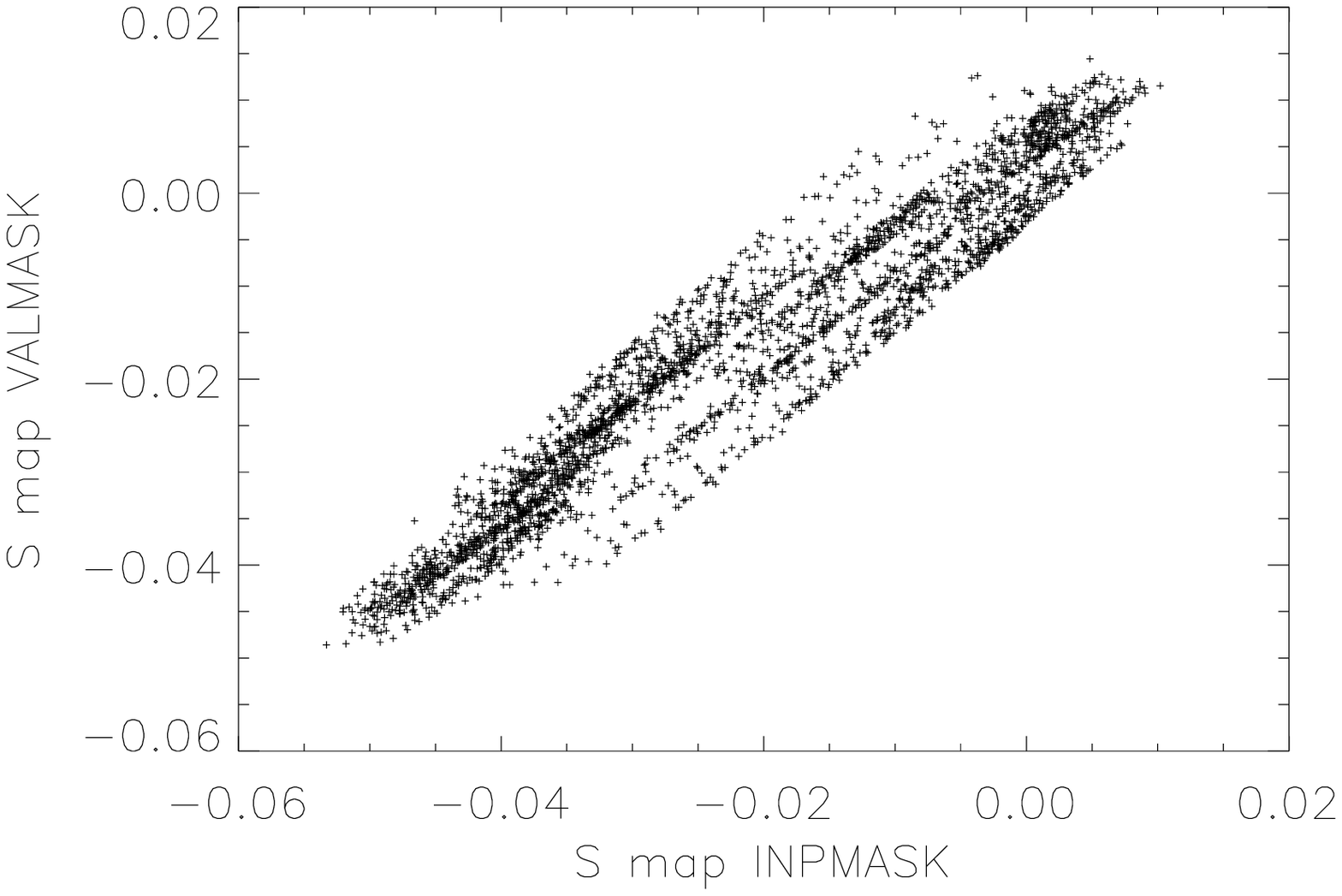}
\hspace{10mm}
\includegraphics[width=6.5cm,height=5.cm]{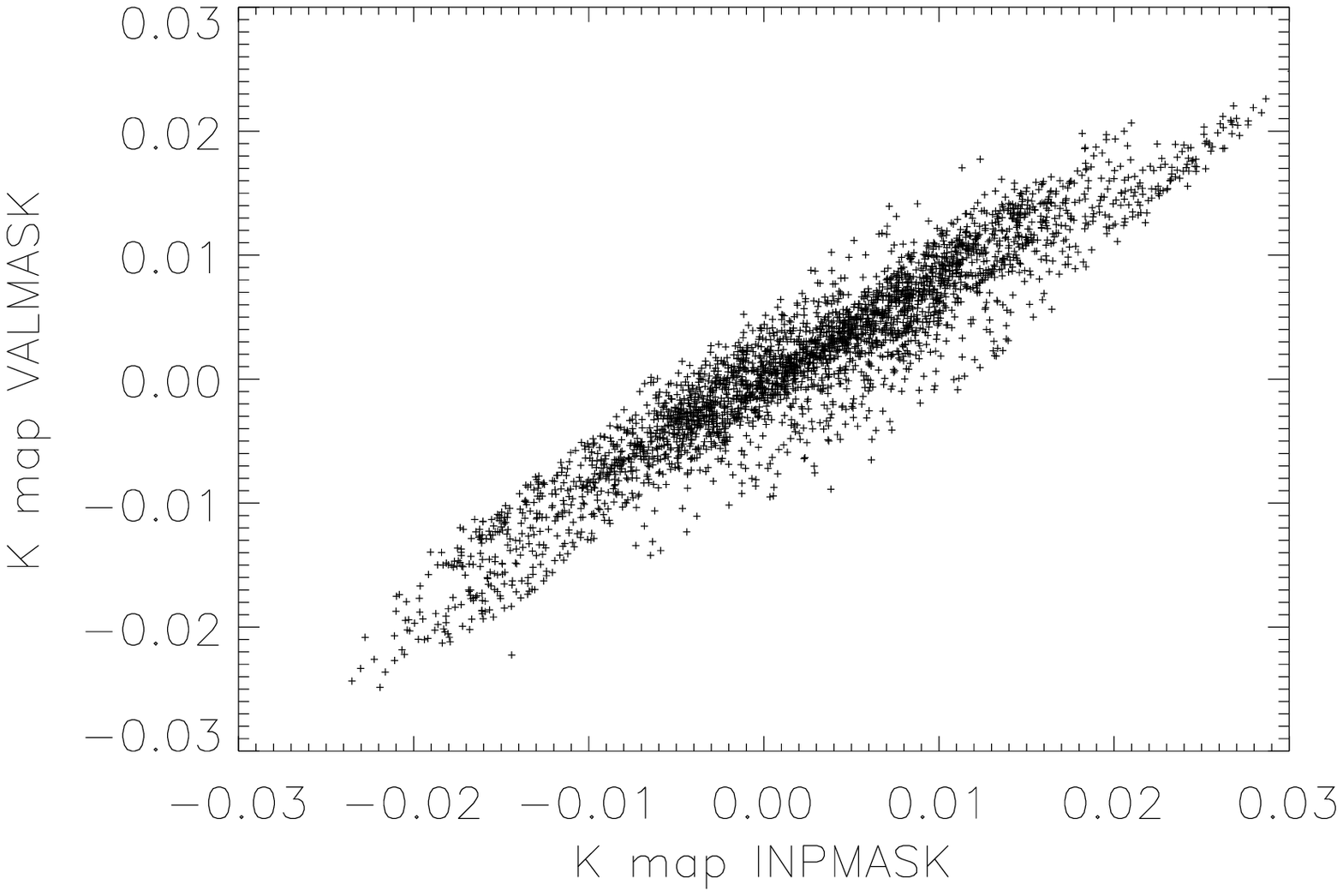}
%
\caption{Correlation between the two $S$ maps (left) and the two $K$ maps (right)
of Figure~\ref{Fig1}. The left panel presents the correlation of two $S$ maps,
one of them masked with the {\sc inpmask}, and another with {\sc valmask}.
The right panel shows the correlation of two $K$ maps with the {\sc inpmask} and
{\sc valmask}. All non-Gaussian maps we employed were generated from CMB input
{\sc smica} map.  \label{Fig2} }
\end{center}
\end{figure*}

Figure~\ref{Fig1} gives an illustration of $S$ (left panels) and $K$ (right panels)
maps obtained from the foreground-cleaned {\sc smica} Planck map with grid resolution
$N_\text{side} = 512$ and $\ell_\text{max} = 500$, which we have used in the analyses
of this paper.
In the first row, the {\sc inpmask}  was used for calculating  
$S$ and $K$ maps, while in the second row the {\sc valmask} was employed to
generate the maps.%
\footnote{We also calculated $S$ and $K$ maps from the {\sc nilc} and {\sc sevem}
maps with the available {\sc inpmask}, {\sc valmask}, and U73 masks.
However, to avoid repetition we only depict the maps of Figure~\ref{Fig1}.}
These maps show distributions of hot (red) and cold (blue) spots (higher and lower values) for the
skewness and kurtosis that are not evenly distributed in the celestial sphere, at first sight suggesting
a large-angle multipole component of non-Gaussianity in the {\sc smica} data, whose statistical
significance we examine in detail below.
A first comparison between the $S$ maps (left panels) and between the  $K$ maps
(right panels) shows several large-scale similarities between the
maps in the columns in Figure~\ref{Fig1}.
To go a step further in this study, in Figure~\ref{Fig2} we show the result of
a pixel-to-pixel comparative analysis of the pairs of $S$ and $K$ maps of Figure~\ref{Fig1}.
Thus, the left hand panel shows the correlation of $S$ maps calculated from {\sc smica}
with {\sc inpmask} and {\sc valmask}, while the right hand panel shows  a similar correlation
of two $K$ maps with these masks.
The panels of Figure~\ref{Fig2} make it apparent that the pair of $S$ and the pair of
$K$ maps,  produced from {\sc smica} and equipped with {\sc inpmask} and {\sc valmask},
are well-correlated, with Pearson's coefficients given by $0.961$ and $0.947$ for
the pair of $S$ maps and the pair $K$ maps, respectively.%
\footnote{At first glance, this could be seen as an indication that Gaussian features of the
{\sc smica} map would not change with the application of {\sc inpmask} and {\sc valmask}.
However, we show below that despite these correlations, the large-scale Gaussianity properties
change appreciably with the mask used.}

To obtain quantitative large angular scale information of the $S$ and $K$ maps obtained from
foreground-cleaned Planck maps, we have calculated the low $\ell$ ($\,\ell=1,\,\,\cdots,10\,$)
power spectra $S_{\ell}$ and $K_{\ell}$  from {\sc nilc}, {\sc smica}, and {\sc sevem} Planck maps
with the released {\sc inpmask}, {\sc valmask}, and U73 masks.
These power spectra can be used to access large-angle (low $\ell$)
deviation from Gaussianity in the Planck maps.
To estimate the statistical significance, we collectively compare $S_\ell$ and $K_\ell$
with the multipole values of the averaged power spectra
$\overline{S}^{G}_{\ell}$ and $\overline{K}^{G}_{\ell}$ calculated
from $1\,000$ Gaussian CMB maps, which have been obtained as stochastic
realizations, with $\Lambda$CDM power spectrum, by using {\sc synfast}
facility of the {\sc healp}ix package (G\'orski et al.~\cite{Gorski-et-al-2005}).
\begin{figure*}[ht!] 
\begin{center}
\includegraphics[width=8cm,height=5.6cm]{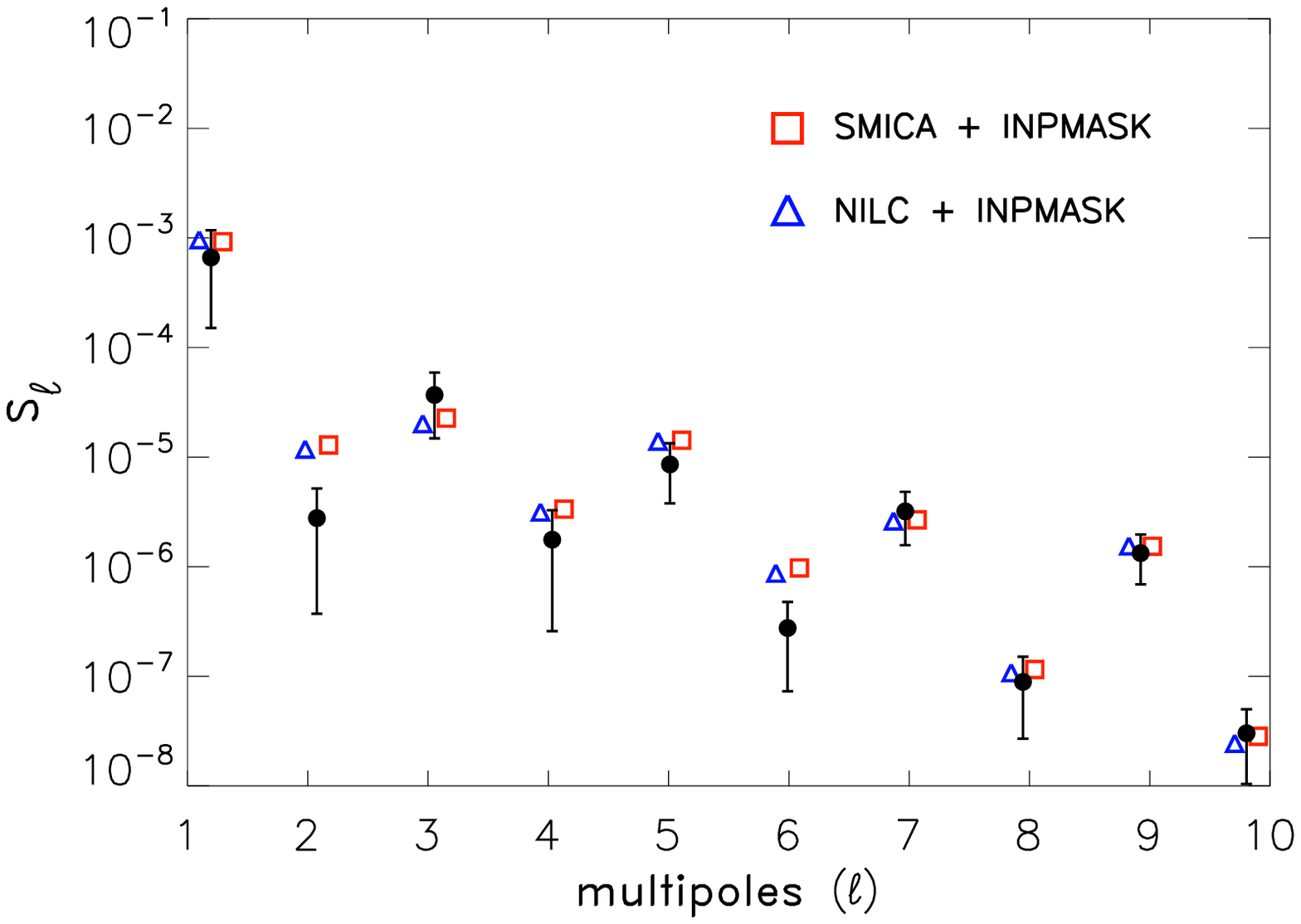}
\includegraphics[width=8cm,height=5.6cm]{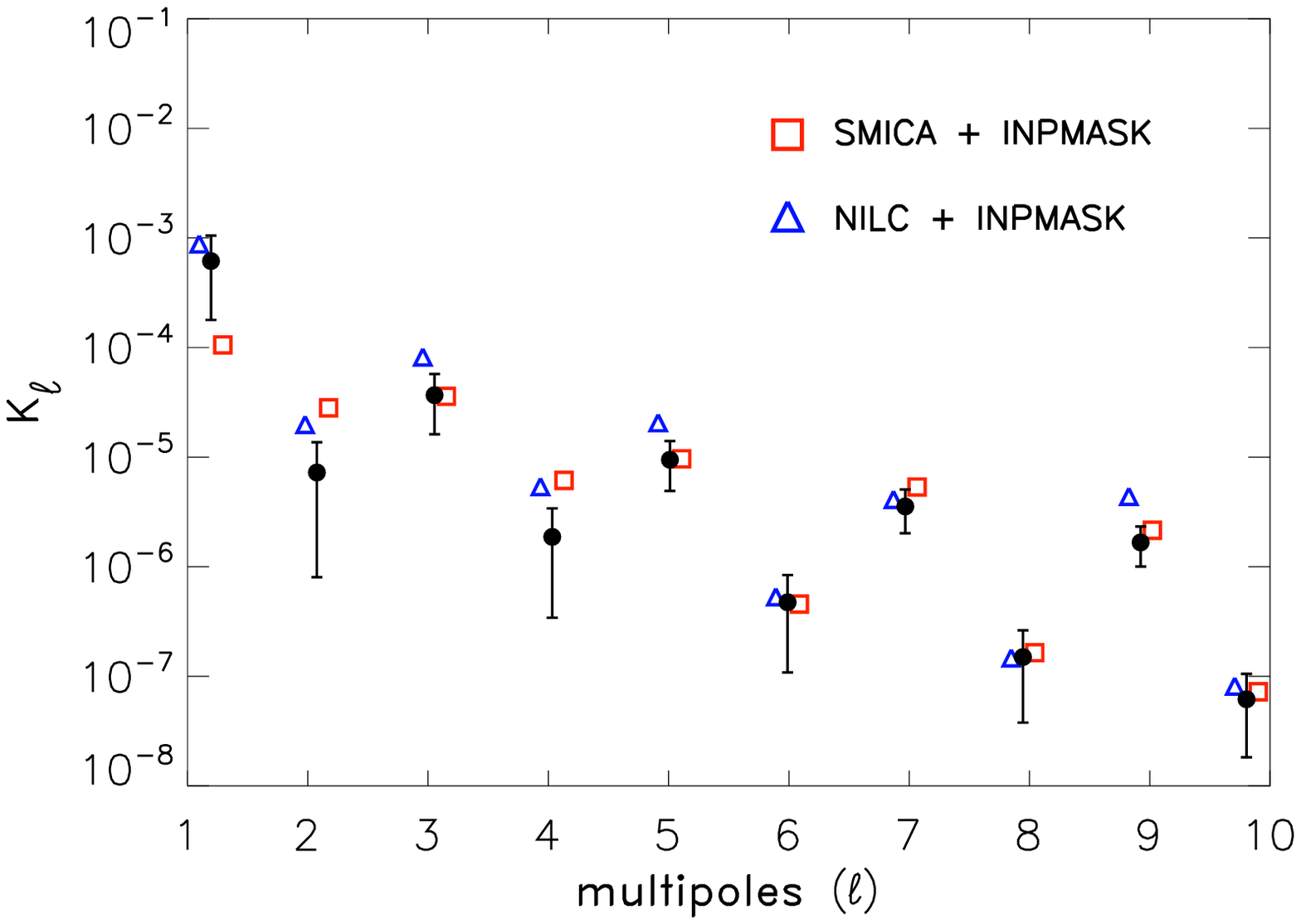}
\includegraphics[width=8cm,height=5.6cm]{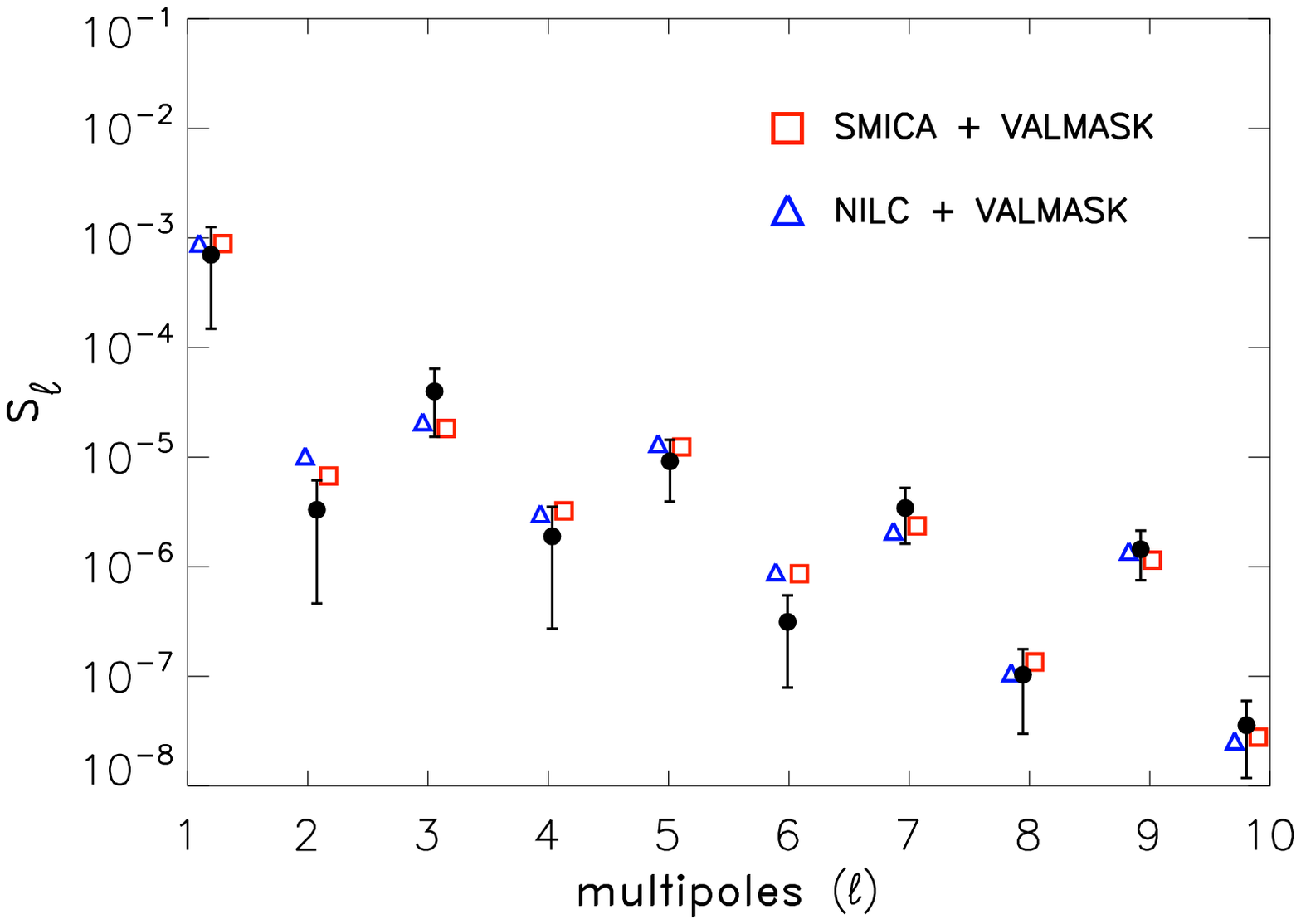}
\includegraphics[width=8cm,height=5.6cm]{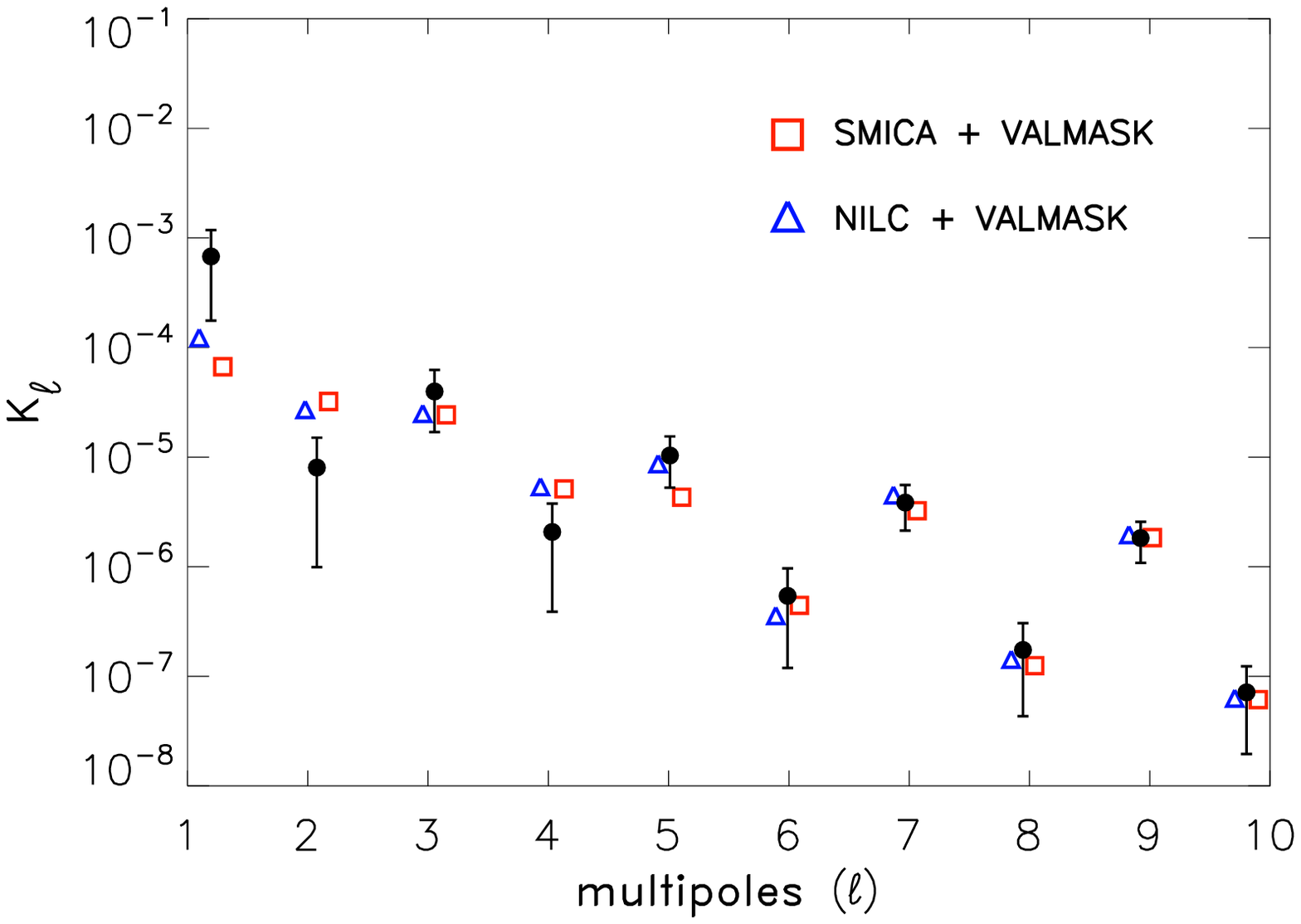}
\caption{Low $\ell$ power spectra $S_{\ell}$ (left panels) and  $K_{\ell}$
(right panels) calculated from {\sc smica} and {\sc nilc} foreground-cleaned
Planck maps equipped with {\sc inpmask} (first row) and {\sc valmask} (second row).
Tiny horizontal shifts have been used to avoid overlaps of symbols.
The $1\sigma$ error bars (68.3 \% confidence level) are calculated
from the power spectra of $1\,000$ Gaussian maps.}
\label{Fig3}
\end{center}
\end{figure*}

\begin{figure*}[thb!] 
\begin{center}
\hspace{-8mm}
\includegraphics[width=8cm,height=5.6cm]{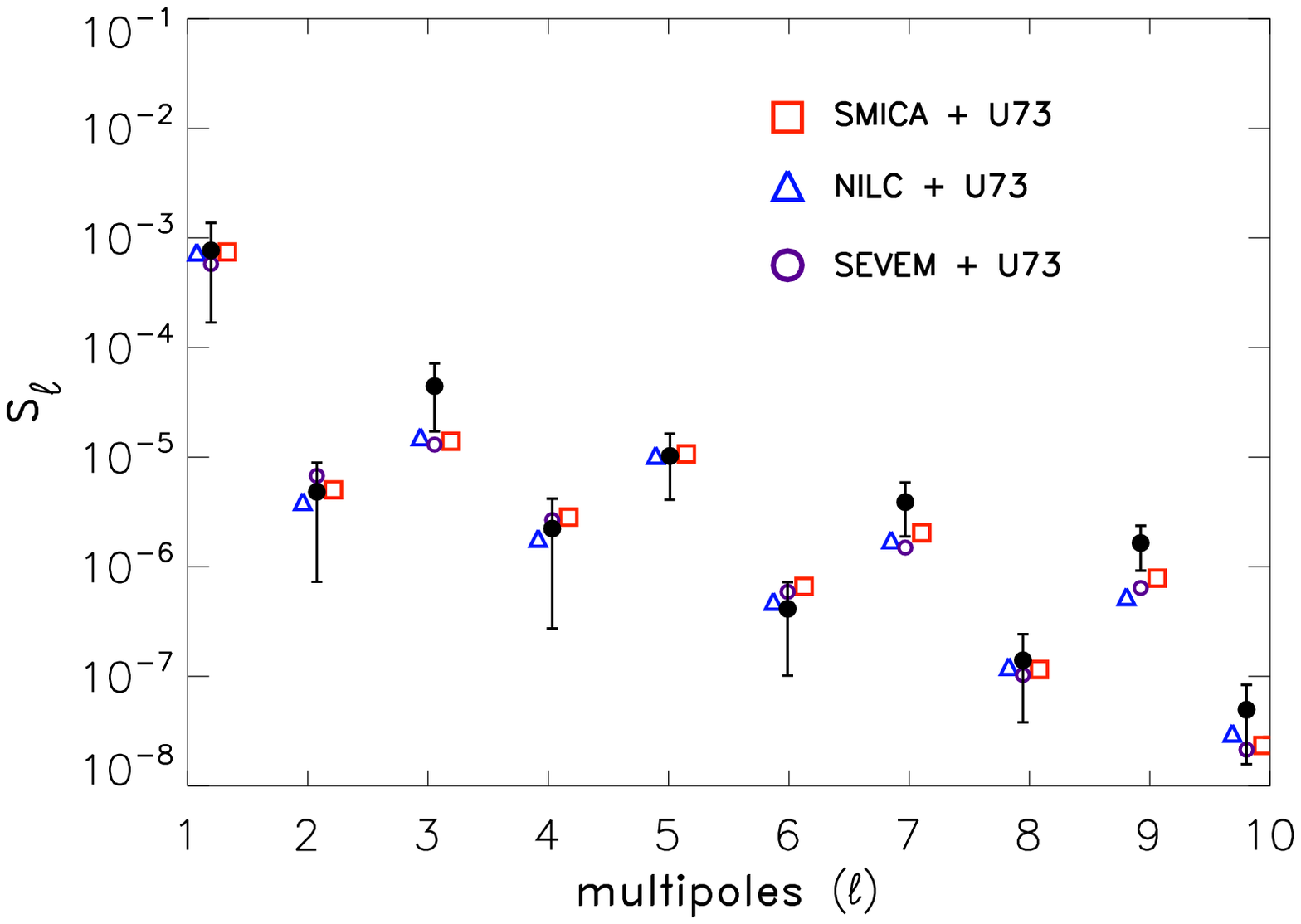}
\hspace{0mm}
\includegraphics[width=8cm,height=5.6cm]{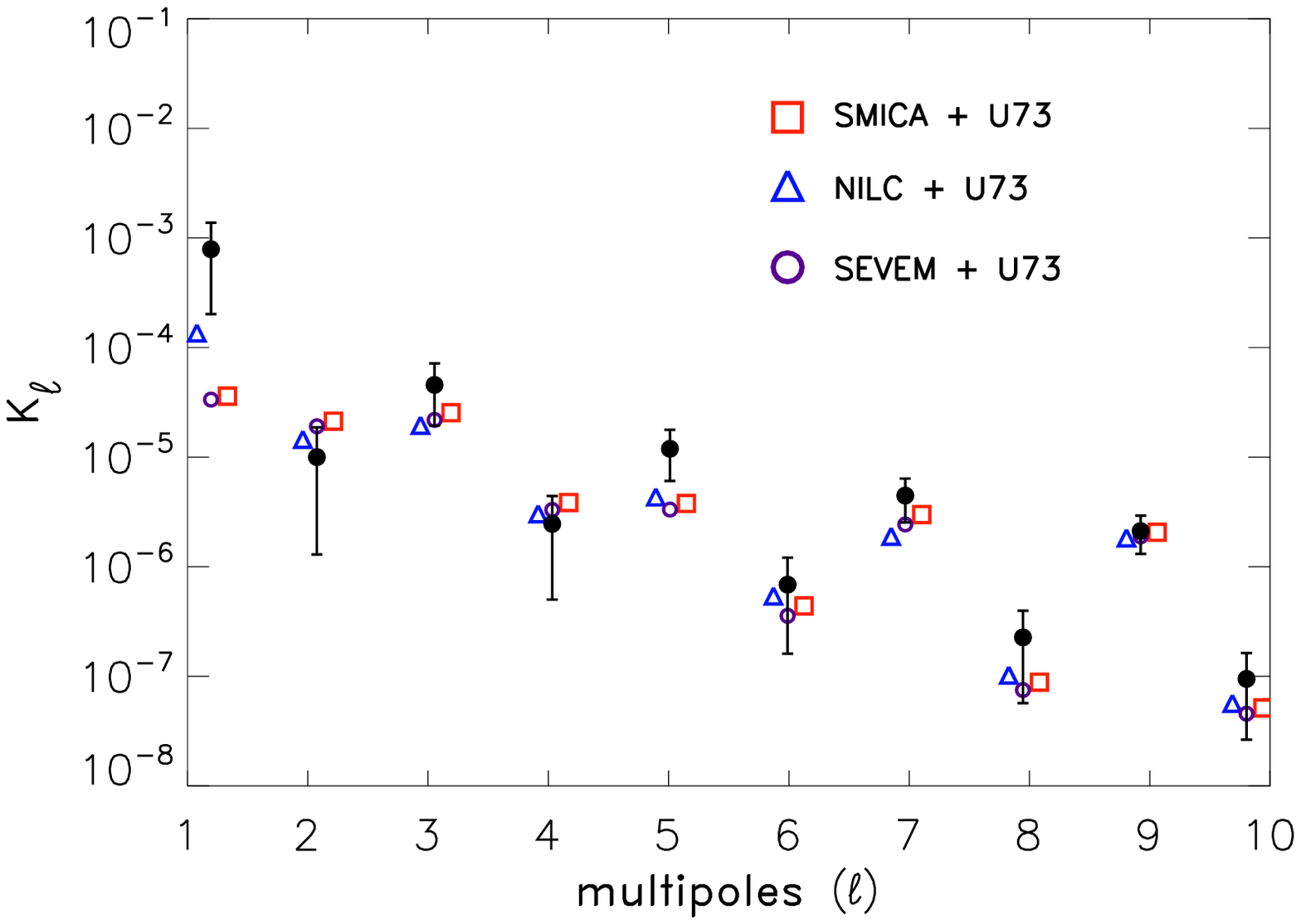}
\caption{Low $\ell$ power spectra $S_{\ell}$ (left) and  $K_{\ell}$
(right) calculated from {\sc smica},  {\sc nilc} and {\sc sevem} foreground-cleaned
U73 masked maps. Tiny horizontal shifts have been used to avoid overlaps of symbols.
The $1\sigma$ error bars (68.3 \% confidence level) are calculated
from the power spectra of $1\,000$ Gaussian maps.}
\label{Fig4}
\end{center}
\end{figure*}

Figure~\ref{Fig3} shows the power spectra $S_{\ell}$  and 
$K_{\ell}$  calculated from  {\sc smica} and {\sc nilc}   
maps masked with {\sc inpmask} (first row) and {\sc valmask} (second row).%
\footnote{We note that there is no available {\sc inpmask} for the
{\sc sevem} map. Thus, we have not included this map in the
analysis of Figure~\ref{Fig3}.}
This figure also displays the points of the averaged power spectra $\overline{S}^{G}_{\ell}$
and $\overline{K}^{G}_{\ell}$ calculated from $1\,000$ Gaussian simulated CMB maps and
the corresponding $1\sigma$ error bars.
To the extent that some of power spectra values $S_{\ell}$ and $K_{\ell}$ fall off the $1\sigma$
error bars centered on $\overline{S}^{G}_{\ell}$ and $\overline{K}^{G}_{\ell}$ values,
this figure suggests deviation from Gaussianity in both  {\sc smica}
and {\sc nilc} maps when masked  either with {\sc inpmask}  or with
{\sc valmask}. This deviation, however, seems to be substantially
smaller when a more severe cut-sky is used. Indeed, in
Figure~\ref{Fig4} it is shown that the power spectra $S_{\ell}$  
and $K_{\ell}$ calculated from {\sc smica}, {\sc nilc},                 
and also {\sc sevem} maps, but now with the U73 mask.
This time, however, all low $\ell$ multipole values are within $1\sigma$ bar of the mean multipoles
values obtained from simulated Gaussian maps.

Although the above comparison of the power spectra can be used as a first indication of deviation
from Gaussianity of different degrees for distinct masks, to have a quantitative overall assessment
of this deviation on a large angular scale, we employed the power spectra
$S_\ell$ and $K_\ell$ (calculated from the  foreground-cleaned Planck maps)
to carry out a $\chi^2$ analysis to determine the goodness of fit of these
power spectra obtained from the Planck maps as compared to the mean power
spectra calculated from simulated Gaussian maps ($\overline{S}^{G}_{\ell}$
and $\overline{K}^{G}_{\ell}$).
Thus,  for the  power spectrum $S_\ell$ obtained from a given Planck
map one has%
\begin{equation}
\chi^2_{S_\ell} = \frac{1}{n-1} \sum_{\ell=1}^{n}\frac{\left({S_\ell}
- {\overline{S}^{G}_{\ell}} \right)^2}{(\,{\sigma_\ell^G}\,)^2}\,,
\label{chi squared}
\end{equation}
where $\overline{S}^{G}_{\ell}$ are the mean multipole values for each $\ell$
mode, $(\sigma_{\ell}^{G})^2 $  is the variance calculated from $1\,000$ Gaussian
simulated maps, and $n$ is the highest multipole one chooses to analyze the
Gaussianity.
We took this to be $\ell = 10$ in this paper, since we are concerned with large-angle non-Gaussianity.
Obviously a similar expression and reasoning can be used for $K_\ell$.

The greater the $\chi^2$ values, the lower the $\chi^2$--probability, that is,
the probability that the multipole values of a given Planck
map $S_{\ell}$ and the mean multipole values $\overline{S}^{G}_{\ell}$ agree.
Thus, for a given Planck map, the $\chi^2_{_{S_\ell}}\!\!$--~probability
measures  its deviation from Gaussianity as detected by the
indicator $S$.
Clearly, the lower the $\chi^2_{_{S_\ell}}\!\!$--~probability, the greater the departure from Gaussianity.
Again, similar procedures and statements hold for $K_{\ell}$ and the corresponding
$\chi^2_{_{K_\ell}}\!\!$--~probability.
In this way, for each foreground-cleaned Planck maps with a given mask,
we can calculate statistical numbers that collectively quantify the
large-angle deviation from Gaussianity as detected by our indicators
$S$ and $K$.

In Table~\ref{Table2} we collect the results of our $\chi^2$ analyses in terms of probability for the
{\sc smica}, {\sc nilc}, and {\sc sevem}, equipped with released {\sc inpmask}, {\sc valmask},
and the union mask U73.
Concerning the {\sc smica} and {\sc nilc}, this table shows significant deviations from Gaussianity
($\, \gtrsim 98\%$ confidence level) for both maps when they are masked with {\sc inpmask}s
(sky-cut is $3\%$), which becomes smaller (with different rate)
when the {\sc valmask}s, whose sky-cuts are, respectively,  $7\%$ and $11\%$,
are employed.%
\footnote{This comparison was not made for the {\sc sevem} map since no {\sc inpmask} for this
map has been released by the Planck team.
We included the values of $\chi^2\!$--~probabilities for the {\sc sevem} masked with {\sc valmask}
in Table~\ref{Table2} for completeness.}
Table~\ref{Table2} also makes clear that, although with different $\chi^2$--~probabilities, the
{\sc smica}, {\sc nilc}, and {\sc sevem} masked with the union mask U73 are consistent with
Gaussianity as detected by our indicator $S$, in agreement with the results found by the Planck
team (Planck Collaboration XXIII~\cite{Planck-XXIII}).
Interestingly, Table~\ref{Table2} shows that this consistency with Gaussianity is less when
the $K$ indicator is employed in the analysis.
This seems to be associated with large-angle anomalies as reported by the Planck
team (Planck Collaboration I; XXIII~\cite{Planck-I}).

In their studies of Gaussianity, the Planck team has performed a large number of validation
tests to examine the impact of realistic factors on their results of non-Gaussianity.
Among these tests they have studied the effect of the different noise models in their estimates and
produced (and released) anisotropic full-sky noise maps for {\sc smica}, {\sc nilc}, and {\sc sevem}
maps (Planck Collaboration XXIV~\cite{Planck-XXIV}).
Thus, a pertinent question that arises here is how the above analyses are impacted if one
incorporates the noise estimated by the Planck team into their foreground-cleaned maps.
To tackle this question, we calculated {\sc smica}+noise, {\sc nilc}+noise, and {\sc sevem}+noise
maps, which were then masked with {\sc inpmask}, {\sc valmask}, and U73 masks, and used the
resulting maps as input in our analyses  of NG, performed through the statistical procedure of
Section~\ref{Indicators}.
In Table~\ref{Table3} we collected the results of our calculations.
The comparison of this with Table~\ref{Table2} shows that the results of the analyses do not change
appreciably when the noise is incorporated into the foreground-cleaned Planck maps.

\begin{table}[!hbt]
\begin{center}
\begin{tabular}{lcc} 
\hline \hline 
Map \ \& \ Mask \ & $\chi^2_{_{S_\ell}}\!\!$--~probability  \ & $\chi^2_{_{K_\ell}}\!\!$--~probability  \\
\hline
{\sc smica}--{\sc inpmask}  &  $1.00\!\times\!10^{-4}$ &  $1.02\!\times\!10^{-2}$ \\
{\sc smica}--{\sc valmask}  &  $3.74\!\times\!10^{-1}$ &  $4.49\!\times\!10^{-2}$ \\
{\sc nilc}--{\sc inpmask}   &  $1.80\!\times\!10^{-3}$ & $2.00\!\times\!10^{-5}$ \\
{\sc nilc}--{\sc valmask}   &  $9.95\!\times\!10^{-2}$ & $1.45\!\times\!10^{-1}$ \\
{\sc sevem}--{\sc valmask}  &  $8.68\!\times\!10^{-1}$ & $5.95\!\times\!10^{-1}$ \\
{\sc smica}--U73            &  $8.43\!\times\!10^{-1}$ & $ 5.14\!\times\!10^{-1}$ \\
{\sc nilc}--U73             &  $8.25\!\times\!10^{-1}$ & $6.27\!\times\!10^{-1}$ \\
{\sc sevem}--U73            &  $7.29\!\times\!10^{-1}$ & $4.56 \!\times\!10^{-1}$ \\
\hline \hline
\end{tabular}
\end{center}
\caption{Results of the $\chi^2$ test  to determine the goodness of fit
for $S_{\ell}$ and $K_{\ell}$ multipole values, calculated from the foreground-cleaned
{\sc smica}, {\sc nilc}, and {\sc sevem} with {\sc inpmask}, {\sc valmask}, and U73 masks,
as compared to the mean  power spectra $\overline{S}^{G}_{\ell}$ and $\overline{K}^{G}_{\ell}$
obtained from ${S}^{G}$ and  ${K}^{G}$ maps generated from $1\,000$ simulated Gaussian
maps masked accordingly.}  \label{Table2}
\end{table}

\begin{table}[!bht]
\begin{center}
\begin{tabular}{lcc} 
\hline \hline 
Map \ \& \ Mask \ & $\chi^2_{_{S_\ell}}\!\!$--~probability  \ & $\chi^2_{_{K_\ell}}\!\!$--~probability  \\
\hline
{\sc smica}+noise--{\sc inpmask}      & $3.00 \!\times\!10^{-4}$ & $1.61\!\times\!10^{-2}$ \\
{\sc smica}+noise--{\sc valmask}      & $4.20\!\times\!10^{-1}$  & $5.19\!\times\!10^{-2}$ \\
{\sc nilc}+noise--{\sc inpmask}       & $1.10\!\times\!10^{-3}$  & $1.00\!\times\!10^{-4}$  \\
{\sc nilc}+noise--{\sc valmask}       & $1.15 \!\times\!10^{-1}$ & $1.62\!\times\!10^{-1}$ \\
{\sc sevem}+noise--{\sc valmask}      & $8.68\!\times\!10^{-1}$  & $6.16\!\times\!10^{-1}$ \\
{\sc smica}+noise--U73                & $8.07\!\times\!10^{-1}$  & $6.47\!\times\!10^{-1}$ \\
{\sc nilc}+noise--U73                 & $8.25\!\times\!10^{-1}$  & $6.01\!\times\!10^{-1}$ \\
{\sc sevem}+noise--U73                & $7.55\!\times\!10^{-1}$  & $5.91 \!\times\!10^{-1}$ \\
\hline \hline
\end{tabular}
\end{center}
\caption{Results of the $\chi^2$ test  to determine the goodness of fit
for $S_{\ell}$ and $K_{\ell}$ multipole values calculated from foreground-cleaned
noise-added masked maps as compared to the values of
the mean  power spectra  $\overline{S}^{G}_{\ell}$ and $\overline{K}^{G}_{\ell}$
obtained from $1\,000$ simulated Gaussian maps masked accordingly.} \label{Table3}
\end{table}

\section{Concluding remarks} \label{Conclusions}

The study of Gaussianity of CMB fluctuations can be used to break the
degeneracy between the inflationary models and to test alternative scenarios.
In most of these studies, one is particularly interested in the primordial component.
However, there are several sources of non-Gaussian contaminants in the CMB data.

One does not expect that a single statistical estimator can be sensitive to all sources
of non-Gaussianity that may be present in observed CMB data.
On the other hand, different statistical indicators can provide valuable complementary information
about different features and be useful for extracting information about the source of non-Gaussianity.
Thus, is it important to test CMB data for Gaussianity by employing different
statistical estimators.

Studies of the non-Gaussianity of the CMB temperature fluctuations data can
be grouped into two classes of statistical approaches. In the first, one searches
to constrain primordial non-Gaussianity parameter such as the amplitude $f_{\rm NL}$,
which can be predicted from the different models of the early universe and confronted
with observations.
Different models give rise to different predictions (type and amplitude) for $f_{\rm NL}$.
Besides allowing an optimal implementation, another advantage of this class of parametric
estimators is that it is easy to compare estimates of $f_{\rm NL}$ for different models.
However, the standard formulation and implementation of the estimator $f_{\rm NL}$ has
been designed under the assumption of statistical isotropy of the CMB sky.
The second class contains statistical tools designed to search for non-Gaussianity in
the CMB maps regardless of its origin.
The surrogate map technique used by Planck team and our $S$ and $K$ indicators are examples
of the statistical procedures in this class.
The indicators of this class have the advantage of being model-independent, since no particular
model of the early universe is assumed in their formulation and practical
implementation.
However, with these model-independent tools, one cannot immediately insure whether a detection of
non-Gaussianity in CMB data is of primordial origin.
This makes the problem of finding the connection between statistical tools in these two classes a difficult
theoretical task.

The Planck team has examined Gaussianity in CMB data by using various different statistical
tools and approaches. Among others, they have employed one of the simplest tests for
Gaussianity of CMB data by calculating a sample of values of skewness $S$ and kurtosis
$K$ from the whole set of accurate temperature fluctuations values of their
foreground-cleaned U73-masked maps, and made a comparison  with the averaged
values for the statistical moments calculated from simulated maps (Planck Collaboration
XXIII~\cite{Planck-XXIII}).

In this work we have gone a step further, and instead of using two dimensionless overall numbers,
we employed a procedure for defining, from a  CMB map, two discrete functions $S(\theta_j, \phi_j)$ and
$K(\theta_j, \phi_j)$, which measure directional deviation from Gaussianity,
for  $j=1, \ldots , 3072$ homogeneously distributed directions, $(\theta_j, \phi_j)$, on the two-sphere
$\mathbb{S}^2$.
The Mollweide projections of these two functions are skewness and kurtosis maps, whose low $\ell$
power spectra has been used to investigate large-angle deviation from Gaussianity in Planck data.
These estimators capture directional information of non-Gaussianity and are useful, for instance,
in the presence of anisotropic signals such as residual foregrounds.

This statistical procedure has been recently proposed (Bernui \& Rebou\c{c}as~\cite{BR2009})
and used in connection with foreground-reduced WMAP maps (Bernui \&
Rebou\c{c}as~\cite{BR2010}; Zhao~\cite{Zhao}).
A significant deviation from Gaussianity has been found in the WMAP foreground-reduced maps,
which varies with the cleaning processes (Hinshaw et al.~\cite{ILC-5yr-Hishaw};
Kim et al.~\cite{HILC-Kim}; Delabrouille et al.~\cite{Delabrouille09}).

This paper addressed several interrelated issues regarding these skewness-kurtosis spherical caps
procedure in connection with large-angle non-Gaussianity in the foreground-cleaned Planck maps.
First, we used these statistical estimators to analyze the Gaussianity of the nearly full-sky
foreground-cleaned Planck maps, therefore extending the results of Ref. Bernui \&
Rebou\c{c}as~(\cite{BR2010}).
Second, we made statistical analyses of the effects of different component separation cleaning
methods in the Gaussianity, and quantified  the level of non-Gaussianity for each foreground-cleaned
Planck maps equipped with each the released {\sc inpmask}, {\sc valmask}, and U73 masks
(see Table~\ref{table1}).
Third, we used the pixel's noise maps released by the Planck team to examine the robustness of
the Gaussianity analyses with respect to this noise.
The main results of our analyses are summarized in Tables~\ref{Table2} and
\ref{Table3}, together with Figs.~\ref{Fig1} to~\ref{Fig4}.

Figure~\ref{Fig1} illustrates skewness (left) and kurtosis (right) maps obtained from  the {\sc smica}
Planck maps with {\sc inpmask} (first row) and {\sc valmask} (second row).
The panels of Figure~\ref{Fig2} illustrate how well the pair of $S$ maps
are correlated, as are the pair of $K$ maps of Fig.~\ref{Fig1}.

The power spectra $S_{\ell}$ and $K_{\ell}$ of the maps in Figure~\ref{Fig3}
indicate deviation from Gaussianity in both  {\sc smica} and {\sc nilc} maps
when masked with either {\sc inpmask} or {\sc valmask}. This deviation,
however, is substantially reduced when a more severe cut-sky is employed
as shown in Figure~\ref{Fig4}, wherein all low $\ell$ multipole values are
within $1\sigma$ bar of the mean multipoles values obtained from simulated
Gaussian maps.

The comparison of Fig.~\ref{Fig3} with Fig.~\ref{Fig4} furnishes rough
indications of deviation from Gaussianity on a large angular scale and
the role played by the different Planck masks.
However, we made quantitative overall assessments of this large-angle deviation through
$\chi^2$ analyses and determined the goodness of fit of the power spectra obtained from the
Planck maps as compared to the mean power spectra calculated from simulated Gaussian maps.

In Table~\ref{Table2} we collected the results of our $\chi^2$ analyses in terms of probability.
Concerning the {\sc smica} and {\sc nilc}, this table shows significant deviations from Gaussianity
for both maps when they are masked with their {\sc inpmask}s (sky-cut is only $3\%$), which
becomes smaller when the {\sc valmask}s, whose sky-cuts are $7\%$ and $11\%$, are employed.
We emphasize that Table~\ref{Table2} also shows that --although to different degrees--
the {\sc smica}, {\sc nilc}, and {\sc sevem} masked with the union mask U73 are consistent with
Gaussianity as detected by our indicator $S$, in agreement with the results obtained by the
Planck team, but through different statistical procedures (Planck Collaboration
XXIII~\cite{Planck-XXIII}).
However, a slightly smaller consistency with Gaussianity has been found when the $K$ indicator is
employed, which seems to be associated with the large-angle anomalies reported by the Planck
team (Planck Collaboration XXIII~\cite{Planck-XXIII}).

We also addressed the question of how the results of Gaussianity analyses are modified when
the noise is added to the foreground-cleaned Planck maps, a point that has not been considered
in Planck Collaboration XXIII~(\cite{Planck-XXIII}).
Tables~\ref{Table2} and~\ref{Table3} show that the results of the analyses do not change
appreciably when the noise is incorporated into the Planck maps.

Finally, we note that by using simulated maps in Bernui \& Rebou\c{c}as~(\cite{BR2012}), we
showed that the indicators $S$ and $K$ do not have enough sensitivity to detect tiny primordial
non-Gaussianity.
This amounts to saying that any significant detection of non-Gaussianity through the $S$ and
$K$ should contain a non-primordial contribution.
This makes clear that these indicators can be used to capture non-Gaussianity components of
non-primordial origin that may be present in foreground-cleaned maps.
Furthermore, $S$ and $K$ indicators have the advantage of providing angular variation (directional
information) of non-Gaussianity of the maps and also deal with the large-angular scale features of
non-Gaussianity.

\begin{acknowledgements}
M.J. Rebou\c{c}as acknowledges the support of FAPERJ under a CNE E-26/102.328/2013 grant.
This work was also supported by Conselho Nacional de Desenvolvimento
Cient\'{\i}fico e Tecnol\'{o}gico (CNPq) - Brasil, under Grant No. 475262/2010-7.
M.J.R. and A.B. thank the CNPq for the grants under which this work was carried out.
We are also grateful to A.F.F. Teixeira for reading the manuscript and indicating the omissions
and misprints.
We acknowledge the use of the Planck data.
Some of the results in this paper were derived using the HEALPix
package (G\'orski et al.~\cite{Gorski-et-al-2005}).
\end{acknowledgements}


\end{document}